\documentclass[preprint,12pt]{aastex}
\usepackage{epsfig}
\usepackage{emulateapj5}
\usepackage{apjfonts}

\newcommand{\chandra}{{\it Chandra}}

\newcommand{\xmm}{{\it XMM-Newton}}

\newcommand{\lum}{\thinspace\hbox{$\hbox{ergs}\thinspace\hbox{s}^{-1}$}}

\newcommand{\sss}{M101 ULX-1}

\makeatletter

\newenvironment{inlinefigure}{%
\def\@captype{figure}%
\noindent\begin{minipage}{0.999\linewidth}\begin{center}}
{\end{center}\end{minipage}\smallskip}

\begin{document}

\def\spose#1{\hbox to 0pt{#1\hss}}
\def\laeq{\mathrel{\spose{\lower 3pt\hbox{$\mathchar"218$}}
     \raise 2.0pt\hbox{$\mathchar"13C$}}}
\def\gaeq{\mathrel{\spose{\lower 3pt\hbox{$\mathchar"218$}}
     \raise 2.0pt\hbox{$\mathchar"13E$}}}

%\slugcomment{Accepted for publication in ApJL}

\title{An Unusual Spectral State of an Ultra-luminous Very Soft X-ray
Source during Outburst}

\author{A.~K.~H.~Kong$^1$, R.~Di\,Stefano$^{2,3}$}
\affil{$^1$ Kavli Institute for Astrophysics and Space Research,
Massachusetts Institute of Technology, 77
Massachusetts Avenue, Cambridge, MA 02139; akong@space.mit.edu}
\affil{$^2$ Harvard-Smithsonian Center for Astrophysics, 60
Garden Street, Cambridge, MA 02138; rd@cfa.harvard.edu}
\affil{$^3$ Department of Physics and Astronomy, Tufts
University, Medford, MA 02155}

\begin{abstract}
We report the results of \chandra\ and \xmm\ observations of a new 
outburst of an ultra-luminous X-ray source in M101.
During a \chandra\ monitoring observation of M101, \sss\ was found to
be in outburst in 2004 December, the second outburst in 2004. The peak 
bolometric luminosity is about $3\times10^{40}$\lum\
($7\times10^{39}$\lum\ in 0.3--7 keV). The outburst spectra
are very soft and can generally be fitted with a
blackbody model with temperatures of 40--80 eV, similar to supersoft X-ray
sources in the
Milky Way and in the Magellanic Clouds. In one
\chandra\ observation, the source spectrum appears to be harder with a
temperature of 150 eV. Such a spectral state is rarely seen in M101
ULX-1 and no X-ray source in the Milky Way shows this kind of spectrum.
However, such an unusual spectral state very likely belongs to a new
class of
X-ray sources, quasisoft X-ray sources, recently discovered in nearby
galaxies. M101 ULX-1 returned to supersoft state in a subsequent
\xmm\ observation.
%This is the first example confirmed by spectroscopy that an
%ultra-luminous source undergoes supersoft/quasisoft state transition.
Based on the two outbursts in 2004, the extremely high luminosity
($L_{bol}=10^{40}-10^{41}$ ergs s$^{-1}$),
very soft X-ray spectra ($kT=40-150$ eV), transient
behavior, and state transition provide strong evidence that M101 ULX-1
harbors an intermediate-mass black hole. 

\end{abstract}


\section{Introduction}

Recent high angular resolution X-ray observations reveal that there are a
large number of ultra-luminous X-ray sources (ULXs) in many nearby
galaxies. ULXs are luminous ($L_X > 10^{39}$
\lum) non-nuclear X-ray point sources with apparent X-ray luminosities
above the Eddington limit for a $\sim 10 M_{\odot}$ black hole. While some
ULXs have been associated with supernovae, many are thought
to be accreting objects with X-ray flux variability observed on
timescales of hours to years.  A
natural possibility is that the compact object is an intermediate-mass
black hole (IMBH) with mass of
$\sim 10^{2-4} M_{\odot}$ (Miller \& Colbert 2004). The origin of
such objects remains uncertain. Some ULXs may
have a stellar-mass black hole with beamed emission
(King et al. 2001; K\"{o}rding et al. 2002).
The majority of ULXs have X-ray emission from 0.1--10 keV. Because the
radius of the event horizon, and therefore the radius of the inner
boundary of the accretion disk increases with black hole mass, the
temperature of the inner disk of IMBHs is smaller than typical stellar
mass black holes. Recent X-ray observations suggest that some ULXs have a
cool accretion disk ($kT\sim 0.1$ keV), suggesting the presence of IMBHs
(Miller et al. 2003,2004; Wang et al. 2004).

In a subset of ULXs which exhibit a soft blackbody component, a few of
them
have very soft spectra with no X-ray emission above 1 keV (Fabbiano et
al. 2003; Di\,Stefano \& Kong 2004; Kong \&
Di\,Stefano 2003; Kong et al. 2004), similar to
supersoft X-ray sources (SSSs) in the Milky Way and the Magellanic
Clouds. The high luminosities
of ultra-luminous SSSs are inconsistent
with typical nuclear burning white dwarf models for Galactic SSSs (e.g.,
van den Heuvel et al. 1992). These
ultra-luminous SSSs could, however, very
well be IMBHs. Their
luminosities and temperatures are consistent with what is predicted for
accreting BHs with masses between roughly 100 and 1000 $M_{\odot}$.
Alternatively, outflows from stellar-mass black holes could also
achieve such a high luminosity and low temperature (King \& Pounds
2003). 

The ULX, M101 ULX-1, is one of the most luminous ULXs. It is very
luminous ($L_{bol} \sim 10^{40-41}$\lum) and has a very soft
X-ray spectrum during outburst (see Kong et al. 2004). The source
underwent an outburst in 2004 December.  
In this Letter, we report a series of \chandra\ and \xmm\ observations
of M101 ULX-1 during the 2004 December outburst. Spectral
fits suggest that the source changed from a supersoft state to
a quasisoft state. We compare these results with previous outbursts
and suggest that M101 ULX-1 may harbor an IMBH.

\section{Outbursts of M101 ULX-1}

M101 ULX-1 was discovered with
{\it ROSAT} and was
confirmed as a SSS with a blackbody temperature of about 100 eV, with {\it
Chandra} (Pence et al. 2001; Mukai et al. 2003;
Di\,Stefano \& Kong 2003). During 2000 March, {\it Chandra} detected it at
$L_X\sim3\times10^{39}$ erg s$^{-1}$, and then in 2000 October, its
luminosity was around $10^{39}$ ergs s$^{-1}$. In 2004, {\it Chandra}
conducted a monitoring program for M101.
Figure 1 shows
the long-term X-ray lightcurve of M101 ULX-1 from 2004 January to
2005 January. Two outbursts occurred in 2004 and M101 ULX-1
was at about $L_X=10^{37}$\lum\ during January, March, May,
September, and November. The X-ray luminosities were about $10^{37}$\lum\ 
and the spectra can be fitted with a power-law model (Kong et al.
2004). The source was found in outburst in 2004 July. The detailed
analysis of the \chandra\ and \xmm\ data was presented in Kong et al.
(2004). The outburst spectra are best described with an
absorbed blackbody model with temperatures of $\sim50-100$ eV; the
peak bolometric luminosity is about $10^{41}$\lum.  Subsequent \xmm\
observations found that M101
ULX-1 returned to the low state at the end of 2004 July (see Fig. 1; Kong
et al. 2004,2005). 

\begin{inlinefigure}
\epsfig{file=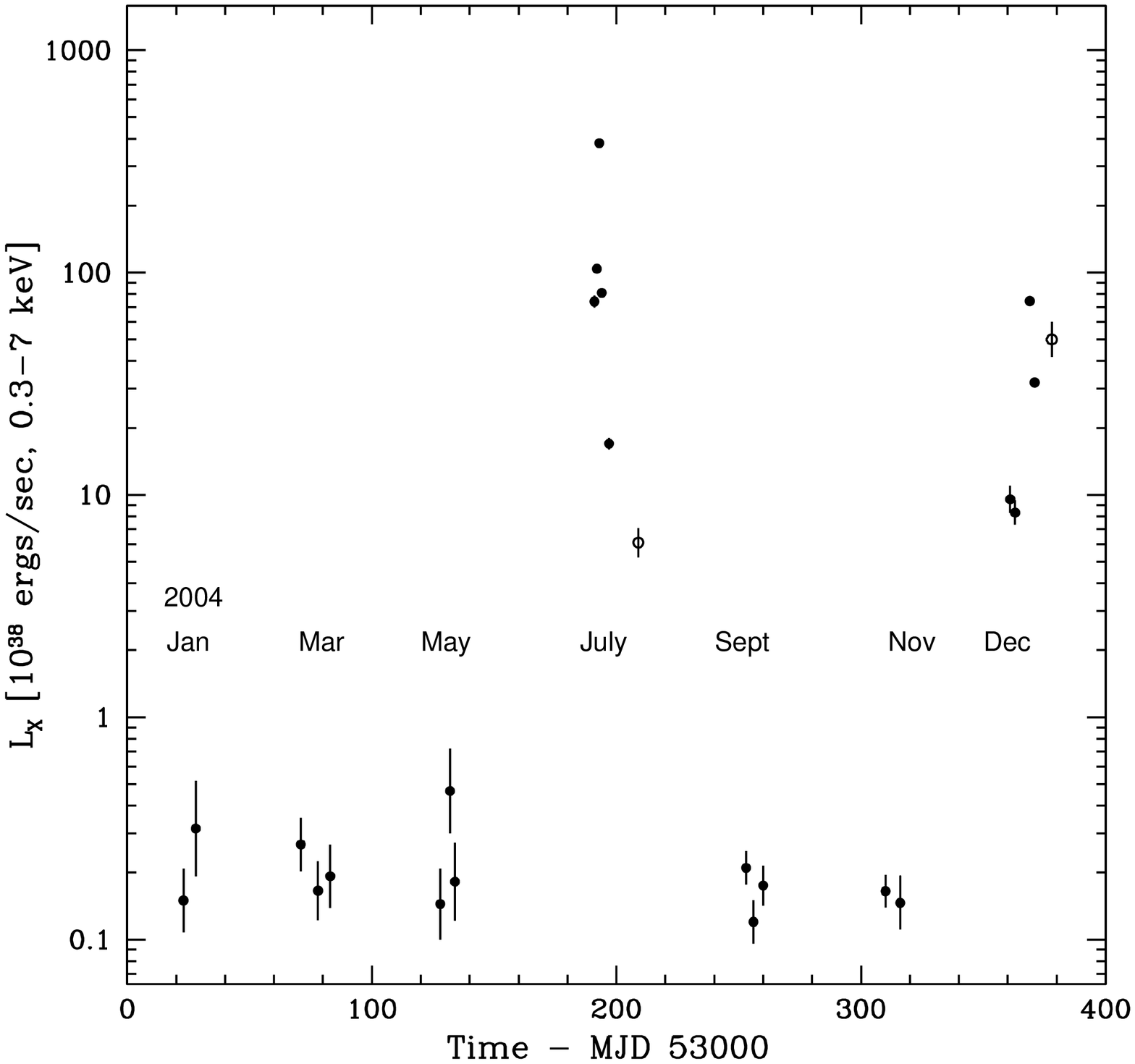,width=3.52in}
\caption{Long-term light curve of M101 ULX-1 observed with \chandra\
  (solid circles) and \xmm\ (open circles) during 2004. The 2
  outbursts are clearly shown. At the time where no data are shown in
  the plot, the source was not observed.}
\end{inlinefigure}

M101 ULX-1 brightened again when \chandra\ observed M101 on 2004
December 22. The count rate jumped from 0.0003 c/s in November to 0.001
c/s on December 22.
Subsequent \chandra\ observations were made on December 24,
30, and 2005 January 1 and the count rate was increasing (see Fig. 1). A
target-of-opportunity \xmm\ follow-up
observation was made on 2005 January 8. No X-ray observations have
occurred since then.

\section{Observations and Data Reduction}

\subsection{\chandra}

M101 was observed with \chandra\ on 2004 December 22, 24, 30, and 2005
January 1. In all observations, the center of M101 and M101 ULX-1 are in
the S3 chip of the ACIS-S but because of the different roll angles, M101
ULX-1 is in different location of the S3 chip in each observation. Each
observation was reprocessed with updated calibration. In
order to reduce 
the instrumental background, we screened the data to allow only photon 
energies in the range of 0.3--7 keV. We also searched for periods of high
background using source free regions in the S1 chip and filtered all the 
high background periods. Only the 2005 January 1 observation 
was affected in which high background lasted for about 1 ks; the exposure
times listed in Table 1 correspond to the good 
time intervals. All data were reduced and analyzed with the
CIAO v3.2.1 package, and calibration database CALDB 3.0.1.

\subsection{\xmm}

Shortly after the discovery of the outburst of \sss\ with 
\chandra, there was an \xmm\ target-of-opportunity follow-up observation.
The \xmm\ observation was taken on 2005 January 8 for 32 ks, and  
the instrument modes were full-frame, thin filter for the three 
European Photon Imaging Cameras (EPIC).
After rejecting those intervals with a high
background level, we considered a good time interval of $17$ ks. Only
data in 0.2--12 keV were used for analysis. We also selected data with
FLAG=0 to reject bad pixels and events too close to chip edges, and 
patterns 0--4 for the EPIC-pn camera and 0--12 for the EPIC-MOS1 and  
EPIC-MOS2 cameras. Data were reduced and analyzed
with the \xmm\ SAS package v5.4.1.
%\footnote{http://xmm.vilspa.esa.es/external/xmm\_sw\_cal/sas\_frame.shtml}.

\section{Analysis and Results}

We performed spectral analysis for all \chandra\ data of the 2004
December outburst shown in Table 1.
The rest of the 2004 data were reported in Kong et al. (2004).
Spectral analysis was performed by using Sherpa. We
also used XSPEC v11.3 for independent check. For models involving pile-up,
we used ISIS for spectral fits.
In each observation, we extracted the source spectrum from a $6''$ radius
circular region 
centered on the source, while an annulus region centered on the source 
was used as the background. Corresponding response files were generated
using CIAO.

During the beginning of the outburst (Dec 22 and 24), count rates were
relatively low; we rebinned
the spectra with at least 5 counts per spectral bin, and applied Gehrels'
approximation (Gehrels 1986) to employ $\chi^2$ statistics to find the
best-fit parameters. For the data taken near the peak of the outburst
(Dec 30 and Jan 1), we grouped the data into spectral bins containing at
least 20 counts and used $\chi^2$ statistics to estimate the best-fit
parameters and their errors.
We tried power-law, 
blackbody, and disk blackbody models with interstellar
absorption. Power-law model yielded a very large photon index ($\sim
10$) in each case and it gave unacceptable fits for the December 30
and January 1 data. Both blackbody and disk blackbody models were
equally acceptable and gave similar fits.
The best-fitting spectral parameters for all \chandra\ 
observations taken during the 2004 December outburst are shown in Table 1.
The inner disk temperatures
are slightly higher than the blackbody temperatures, but the normalizations
(hence the inner disk radii) often have large error ranges. In general, the 
blackbody temperatures are between $\sim 40-80$ eV except for the
January 1 data.
$N_H$ is 
about $(1-5)\times10^{21}$ cm$^{-2}$, significantly higher than the
Galactic value. 
The 0.3--7 keV unabsorbed luminosity ranges from $10^{39}$\lum\ to
$10^{40}$\lum. The peak bolometric luminosity derived from
blackbody fit on December 30 is about
$2\times10^{40}$\lum. 
It is worth noting that the January 1 data suffered
from pile-up ($\sim 12\%$) due to the high count rate. We included
pile-up effect when fitting the data using the pile-up model in ISIS
(Davis 2001).
Clearly, M101 ULX-1 became significantly hotter ($kT_{BB}=149$ eV) on
2005 January 1; the bolometric luminosity is $4\times10^{39}$ ergs 
s$^{-1}$. Figure 2 shows the
blackbody fits of the 2004 December 30 and 2005 January 1 data.

We used similar techniques to analyze the \xmm\ data. We extracted the
source spectrum from a $15''$ radius circle centered on the source, and
the background spectrum from a source free region.
Response files were generated for each detector using SAS tools. Prior to
the spectral fit, the spectra were grouped to have at least 20 counts per
bin.  M101 ULX-1 was brighter than the January 1 \chandra\ observation
with about 
1200 background-subtracted counts in EPIC-pn. 
We fitted the EPIC-pn, EPIC-MOS1, EPIC-MOS2 spectra simultaneously with
several single-component models. Like the \chandra\ data, only
blackbody and disk blackbody models gave acceptable fits (see Table 1).  
The 0.3--7 keV luminosity is about
$5\times10^{39}$\lum while the bolometric luminosity is
$3\times10^{40}$\lum, suggesting that the source was rebrightening or in
the plateau phase. The best
fitting blackbody fits of the \xmm\ spectra
are shown in Figure 2.

\vspace{5mm}
\begin{inlinefigure}
\rotatebox{-90}{\epsfig{file=f2a.eps,width=2.2in}}
\rotatebox{-90}{\epsfig{file=f2b.eps,width=2.2in}}
\rotatebox{-90}{\epsfig{file=f2c.eps,width=2.2in}}
\caption{\chandra\ spectra of M101 ULX-1 taken on 2004 December 30 (top)
and 2005 January 1 (middle). The spectra can be described by a blackbody
model with $kT=70$ eV (December 30) and $kT=149$ eV (January 1). The
bottom panel shows the \xmm\ spectra taken on 2005 January 8. The
spectral fit is a blackbody model with $kT=55$ eV. MOS1, MOS2, and pn
data are marked as open squares, open triangles, and filled circles,
respectively.
  }
\end{inlinefigure}

\section{Discussion}

M101 ULX-1 is a transient ultra-luminous SSS and it underwent a
second outburst in 2004. The outburst spectra became supersoft ($kT< 100$
eV) but 
the spectrum of 2005 January 1 data is clearly different, suggesting a
spectral change during outburst. Majority of the
outburst spectra are supersoft with no significant X-ray
emission above 1 keV, while the January 1 spectrum is hotter with
$kT_{BB}=149$ eV ($kT_{in}=175$ eV) and a high energy tail can be seen up
to 5 keV. 
There are no known X-ray sources in the universe that show such an
unusual
spectral transition but the spectral behaviors can be associated with a
new class
of X-ray sources, quasisoft X-ray sources (QSSs), recently discovered in
nearby galaxies (Di\,Stefano \& Kong 2004). 
The signature of
QSS is a spectrum with 100 eV $< kT < 350$ eV and $L_X > 10^{36}$ ergs
s$^{-1}$, which is consistent with
the one seen in January 1. In fact, if we follow the classification procedure
described in Di\,Stefano \& Kong (2004), M101 ULX-1 belongs to SSS in
all outburst observations except for January 1 during which the source is
classified as a QSS.  Moreover, the quasisoft state of M101 ULX-1 is
ultra-luminous ($L_{bol} = 4\times 10^{39}$ ergs s$^{-1}$).
To summarize the 2004 December outburst of M101 ULX-1, the source
changed from a SSS to a QSS when it was near the peak of the outburst,
and then became a SSS
again. The supersoft and quasisoft state
transition strongly suggests that SSSs
and QSSs are closely related.
Indeed, as discussed in
Di\,Stefano \& Kong (2004), the distinction between
SSSs and QSSs is phenomenological and not physical. They may represent the
same class of sources. Nevertheless, QSSs appear to represent states that
have not yet
been well studied. In fact, there are no known examples of these sources
in the Milky Way or the Magellanic Clouds. 

\begin{table*}
\centering{
\footnotesize
\caption{Best-fitting Spectral Parameters}
\begin{tabular}{lcccccccccccccc}
\hline
\hline
 &  & &\multicolumn{5}{c}{Blackbody} & & \multicolumn{5}{c}{Disk
Blackbody}
\\
 \cline{4-8} \cline{10-15}\\ 
  & Exposure& Counts& $N_H$$^a$ & $kT_{BB}$  & $\chi^2_{\nu}/$dof &
$L_X$$^a$ & $L_{bol}$$^b$& &$N_H$ &
$kT_{in}$ & norm$^c$ & $\chi^2_{\nu}/$dof & $L_X$$^a$ \\
 Date & (ks) & &($\times10^{21}$ cm$^{-2}$) & (eV) & & & & &
($\times10^{21}$
 cm$^{-2}$) & (eV) &  & &\\
\hline
Dec 22 & 48 & 44&$< 5.5$ & $45^{+23}_{-23}$ & $0.47/5$ & 9.6 &85 &&&
\hspace*{-3.1cm}$2.0^{+1.2}_{-0.8}$ & \hspace*{-2.8cm}$52^{+35}_{-8}$ &
\hspace*{-3.2cm}$19764^{+\infty}_{-19347}$
& \hspace*{-2.1cm}$0.48/5$ & \hspace*{-0.9cm}7.0\\
Dec 24 & 41 & 55&$< 4.5$ & $42^{+24}_{-24}$ & $0.5/6$ &
8.3 & 102&& $1.5^{+1.0}_{-0.6}$ & $47^{+4}_{-6}$ &
$49109^{+\infty}_{-47836}$ & $0.5/6$ & 8.3 \\
Dec 30 & 29 & 618&$2.2^{+0.8}_{-0.8}$ & $70^{+7}_{-7}$  & $1.1/20$ &
74.4 & 196&& $2.6^{+0.2}_{-0.2}$ & $77^{+8}_{-6}$ &
$18896^{+82233}_{-13934}$ & $1.1/20$ & 122 \\
Jan 1 & 20 & 1487&$0.9^{+0.1}_{-0.1}$ & $149^{+3}_{-3}$ & $1.0/53$ &
32&39& & $1.5^{+0.1}_{-0.1}$ & $175^{+1}_{-2}$ &
$94^{+4}_{-5}$ & $0.96/53$ & 54 \\ 
Jan 8 & 17 & 1699&$2.1^{+0.5}_{-0.4}$ & $55^{+3.9}_{-3.7}$ & 0.93/61 &
53.7&300 &&
$2.1^{+0.1}_{-0.1}$ & $63^{+0.8}_{-0.8}$ & $58860^{+5899}_{-5340}$ &
0.96/61 & 70\\ 
\hline
\end{tabular}
}
\par
\medskip
%\hspace{0.005cm}
\begin{minipage}{0.94\linewidth}
\footnotesize
NOTES --- All quoted uncertainties are 90\% confidence. The
spectral fits of January 1 data are corrected for pile-up.\\
%$^a$ in units of $10^{21}$ cm$^{-2}$.\\
$^a$ 0.3--7 keV unabsorbed luminosity ($\times 10^{38}$\lum), assuming 6.7
Mpc (Freedman et al. 2001).\\
$^b$ Bolometric luminosity ($\times 10^{38}$\lum).\\
$^c$ in units of $\left(\frac{R_{in}}{D_{10}}\right)^2 \cos\theta$, where
$R_{in}$ is
the inner disk radius (in km), $D_{10}$ the distance (in 10 kpc) to the
source, and $\theta$ the angle of the disk.

\end{minipage}
\end{table*}

In theory, luminous SSSs and QSSs are good candidates for IMBHs because of
their high luminosities and low temperatures
(Di\,Stefano \& Kong 2003,2004). However, there are only very few examples
that X-ray spectroscopy can be done due to the low count rates and we
cannot conclusively determine whether they are black hole candidates. M101
ULX-1 is a very unique source that provides strong evidences for an IMBH
accretor.  As discussed in Kong et al. (2004), white dwarf and neutron
star models are very unlikely because there is no obvious explanation
for the extreme luminosity (Kong et al. 2004).
There is the suggestion that the black hole accretor of M101 ULX-1 is
indeed a
stellar mass
object with outflow (King \& Pounds 2003; Mukai et al. 2003). However,
this model cannot explain the changes of
temperature, the extremely high luminosities, and the state transition in
the recent outbursts.
Moreover, such a stellar-mass black hole model is based on spectral
analysis with piled-up
spectra during the 2000 March outburst (Mukai et al. 2003). It is
worth noting that without proper
care of a piled-up spectrum, the resulting spectral fit will become harder
than the true value and the luminosity will also be estimated incorrectly 
(Davis 2001). In fact, Mukai et al. (2003) noticed that there was an
excess
in high energies and they restricted the fit below 1.5 keV. While they
also mentioned that an additional power-law may explain the excess, we
suspect that the excess is due to pile-up.
For a direct comparison to the results reported
here, we followed the procedures outlined in Mukai et al. (2003) and
re-analyzed
the 2000 March outburst data with a pile-up model. While the spectral fits
during low and medium luminosity states are not significantly affected by
pile-up, the high state spectrum has about 10\% pile-up. We also performed
a MARX simulation and confirmed that at this high count rate level,
pile-up is unavoidable. 

We first fit the 0.3--1.5 keV
spectrum without a pile-up model and the results are consistent
with Mukai et al. (2003; $N_H=2.8\times10^{20}$ cm$^{-2}$, $kT_{BB}=0.17$
keV, $L_{bol}=3\times10^{39}$ ergs s$^{-1}$, $\chi^2_{\nu}=1.1$). 
We then include pile-up in the
spectral fit
(0.3--7 keV) using the pile-up model in ISIS developed by Davis (2001).
The resulting blackbody temperature drops from 0.17
keV to 0.14 keV while the $N_H$ raises to $5.5\times10^{20}$ cm$^{-2}$;
the bolometric luminosity is about $6.3\times10^{39}$
ergs s$^{-1}$, a factor of 2 greater than Mukai et al. (2003). The
argument for a stellar mass black hole accretor is that the bolometric
luminosities during the outburst are
roughly at $3\times10^{39}$\lum, suggesting an Eddington limit for a
$10 M_{\odot}$ black hole.
Clearly, 
the bolometric luminosity is no longer a constant and the finding is
inconsistent with
the suggestion that M101 ULX-1 is a stellar-mass black hole with outflows,
accreting near the Eddington limit. However, if the stellar-mass black 
hole has a variable beaming factor, it may explain the luminosity 
variation (Fabbiano et al. 2003). Nonetheless, the 2000
March outburst also shows a SSS/QSS transition; the spectrum of the first
30 ks (the flaring state) is quasisoft ($kT=0.14$ keV) while it returns to
supersoft ($kT=0.09$ keV) later.

Based on the observed temperature ($kT < 100$ eV)
and bolometric luminosity ($L_{bol}\sim 10^{40-41}$ ergs s$^{-1}$) during
the 2004 July
outburst, the black hole mass of M101
ULX-1 is estimated to be greater than 2800 $M_{\odot}$ (Kong et al. 2004).
Using the 90\% lower limits of the disk blackbody fits derived from the
2004 December outburst, the black hole mass is
$> 1300-3\times10^4 M_{\odot}$. In particular, the quasisoft spectrum is
consistent with the cool disk model for ULXs that has also been
considered as evidence for IMBHs (Miller et al. 2003,2004; Wang et al.
2004). It is, however, worth noting that the cool disk model for ULXs
usually has a
power-law component which contributes a significant fraction of X-ray
emission and it suggests a corona around the accretion disk. The lack of
power-law component of M101 ULX-1 may simply imply that the disk emission
is not Comptonized effectively or the corona does not exist. In contrast,
IC 342 X--1 appears to have the disk
emission completely Comptonized so that it can be characterized by a
power-law alone (Kong 2003; Wang et al. 2004). 

We expect that at least
some of the ULXs
involving IMBH accretors
are transients (King et al. 2001; Kalogera et al. 2004). The required
condition is that the donor must be a massive star ($\gaeq 5 M_{\odot}$)
in regions of young populations.  More recently, an
optical counterpart of M101 ULX-1 was identified (Kuntz et al. 2005; Kong
et al. 2005) from Hubble Space Telescope images. The colors and spectrum
of the optical counterpart are consistent with a B supergiant with a mass
of 9--12 $M_{\odot}$ (Kuntz et al. 2005). The source is also very close to
star forming regions in a spiral arm (Kong et al. 2005). These results
suggest that the stellar environment around M101 ULX-1 is suitable to
house an IMBH.

In summary, we detected an outburst from M101 ULX-1 and this is the second
outburst in 2004. During the outburst, the peak bolometric luminosity is
about $2\times10^{40}$ \lum\ and the spectra can be fitted with either a
blackbody model or a disk blackbody model. The spectrum is supersoft
except
for the January 1 data during which the spectrum is consistent with QSSs
in other external galaxies. 
Combining the fact that M101 ULX-1 is close to
star forming region and has an optical counterpart consistent with a B
supergiant, the compact object of M101 ULX-1 is likely to be an IMBH.

\begin{acknowledgements}
We thank an anonymous referee and John Davis for the discussion of
pile-up.
This work is supported by NASA grant NNG04GP58G.
This work is based on observations obtained with \xmm, an ESA mission with
instruments and contributions directly funded by ESA member states and
the US (NASA). 
\end{acknowledgements}

\end{document}